# Using ChatGPT in HCI Research—A Trioethnography

Smit Desai

School of Information Sciences, University of Illinois at Urbana Champaign, smitad2@illinois.edu

Tanusree Sharma

Informatics, University of Illinois at Urbana Champaign, tsharma6@illinois.edu

Pratyasha Saha

Department of Physics, University of Dhaka, pratyasha.saha1195@gmail.com

This paper explores the lived experience of using ChatGPT in HCI research through a month-long trioethnography. Our approach combines the expertise of three HCI researchers with diverse research interests to reflect on our daily experience of living and working with ChatGPT. Our findings are presented as three provocations grounded in our collective experiences and HCI theories. Specifically, we examine (1) the emotional impact of using ChatGPT, with a focus on frustration and embarrassment, (2) the absence of accountability and consideration of future implications in design and raise (3) questions around bias from a Global South perspective. Our work aims to inspire critical discussions about utilizing ChatGPT in HCI research and advance equitable and inclusive technological development.

CCS CONCEPTS •**Human-centered computing ~ Human computer interaction (HCI) ~ Interaction paradigms ~ Natural language interfaces**

**Additional Keywords and Phrases:** Trioethnography, ChatGPT, Large Language Models (LLMs), Situated XAI

## 1 INTRODUCTION

Large Language Models (LLMs) have demonstrated remarkable success in a diverse array of natural language tasks, such as machine translation, question answering, and automatic summarization, owing to their state-of-the-art transformer architecture and two-stage training pipeline [55]. The transformer architecture enables LLMs to understand complex relationships between input elements, while their two-stage training process allows them to leverage knowledge acquired during pretraining on large amounts of unannotated data. Among the most prominent LLMs is the ChatGPT [62], an OpenAI-developed conversational artificial intelligence system, boasting more than 175 billion parameters and possessing a multitude of advanced capabilities.

While ChatGPT attracted the attention of researchers and various stakeholders for its potential applications in various fields, the use of ChatGPT in education, research, and healthcare for tasks such as scientific writing, optimizing healthcare workflows, and augmenting personalized learning, necessitates a prudent approach. Several studies indicated concerns regarding the potential limitations and biases, including ethical, transparency, legal, as well as risk of bias [30]. For example, concerns regarding ChatGPT use in healthcare include inaccurate content, cybersecurity issues, and the risk of infodemic [8]. Similarly, the possibility of ChatGPT's misuse in medicine and research may accelerate the production of fake evidence and materials that have a high level of plausibility, leading to fraudulent use. As such, the future use of ChatGPT across diverse domains is a subject of current debates, and it is crucial to critically evaluate limitations and potential biases.

Despite their meteoric rise in popularity and their widespread adoption in a range of domains, there is a lack of qualitative studies exploring the experiential use of LLMs by experts. Addressing this critical gap in the literature, we conducted a rigorous trioethnography to elucidate our own experiences with ChatGPT. Drawing on our collective insights, we endeavor to chart a new trajectory for the use of ChatGPT in Human-Computer Interaction (HCI) research, articulating a set of thought-provoking implications that have the potential to catalyze further innovation in this rapidly evolving field.

## 2 BACKGROUND

In HCI, there is a growing emphasis on using first-person methods to gain a deeper understanding of the role of technology in everyday life by incorporating the cultural meanings of technology use in context [35]. One such method is autoethnography, which has gained significant attention as a valuable approach to comprehending how technology is utilized in daily life from the perspectives of both developers and researchers. By utilizing lived experiences, autoethnography provides a nuanced and insightful examination of the situated use of technology (e.g., [9, 28, 33, 34, 36]).

In contrast to autoethnography, duoethnography and trioethnography focus on the "dialogical" relationships between researchers, with the goal of juxtaposing their experiences to find similarities and differences and construct meaning based on shared realities [48]. We chose trioethnography because we wanted to bring together early-stage researchers with different kinds of HCI expertise and cultural backgrounds to understand the use of ChatGPT. This enabled us to connect as researchers and confront uncomfortable emotions and experiences that might have otherwise gone unexplored or unacknowledged.

## 3 TRIOETHNOGRAPHIC PROCESS & POSITIONALITY

Our approach to the trioethnogrpahy is inspired by [27]. To initiate the project, Smit, with expertise in conversational AI, sought to form a team that could offer diverse viewpoints on using LLMs in HCI research. The goal was to write a provocation paper for CUI 2023. Smit connected with Tanusree on social media because of her experience in usable security and governance tooling and shared interest in LLMs. The two researchers then contacted Pratyasha, who specializes in HCI for development (HCI4D) and Explainable AI (XAI), to bring a different perspective to the team.

We began our trioethnography on March 6th, 2023. The scope of the trioethnography was limited to using ChatGPT for HCI research or work-related purposes. All other interactions were outside the scope of this study and were not to be recorded. We used a shared Google Doc to journal all our interactions with ChatGPT using annotated screenshots, notes, and reflections. To discuss and juxtapose our experiences, we commented on each other's notes and met weekly for reflective discussions. We recorded and transcribed these meetings and used the transcriptions to write memos about what stood out. We concluded our data collection on April 6th, 2023. Our collective notes, memos, screenshots, and reflections amounted to 10,127 words and 95 pages. We utilized ChatGPT for a variety of purposes, including writing, mundane tasks (such as transcription and image description), information retrieval, and coding.

To ground our trioethnography, we share our positionality. Smit was born in Urban India and migrated to the U.S. for a doctorate. His research prioritizes designing accessible conversational interfaces for older adults. Tanusree was born in rural Bangladesh and migrated to the U.S. for a doctorate. Her research involves building frameworks and user-facing tooling for emerging technologies. Pratyasha was raised in and works in Bangladesh as a postgraduate researcher studying social justice, sustainability, and policy design for the Global South. All three authors are HCI researchers with experience in publishing at SIGCHI venues and identify as cisgender BIPOC.

## 4 PROVOCATIONS

As is typical in trioethnographies, we present our individual reflections and direct quotes in a first-person narrative and supplement our analysis by integrating relevant HCI literature to contextualize our experiences and stimulate discussions. The provocations presented in this section serve to accentuate the divergent and unique perspectives of the authors and should not be construed as a unified or synthesized viewpoint. Rather, they are intended as a conflation of discussion and findings to incite further contemplation, inquiry, and, ideally, debates. After studying the primary data, each author proposed several provocations in a meeting after the end of data collection. We selected three based on their ability to spark discussions and ideas among us, their potential relevance to the CUI community, and the theme of CUI 2023 of designing inclusive conversation.

### 4.1 Cold as Ice, Sharp as Knife: The Emotional Paradox of ChatGPT

*Smit: After writing a review for a conference paper, I felt some of my critiques could sound a bit harsh. So, I decided to check with ChatGPT if my review was rude. To my chagrin, ChatGPT called my review "insulting." I did not think it was insulting (at all). I asked ChatGPT to make my review less rude and more constructive. Its output did change the tone of the review but adulterated the meaning. The ChatGPT revised review sounded generic and unactionable. When I told this to ChatGPT, it apologized. But that did not matter as I felt equally embarrassed and frustrated. I decided to submit my original review with some minor edits, but this interaction left a sour taste. Nevertheless, after that, I wrote three more reviews and, despite my previous experience, used ChatGPT to assess the rudeness of them all. I felt I was seeking ChatGPT's approval, even though I understood its limitations.*

In response to this reflection, Tanusree and Pratyasha echoed the sentiment and discussed how embarrassment could create friction between a user and ChatGPT. Although notoriously understudied in HCI [15], it is hypothesized that embarrassment stems from the heightened perceived social presence of the 'other' and is considered an anthropomorphizing behavior [10, 12, 42], typically experienced in public settings [13]. Despite all the unfounded hype [45] and fallacious claims [53], all three of us agreed that ChatGPT has no agency—nonetheless, embarrassment persisted. Interestingly, Smit continued using ChatGPT to assess the tone of his reviews. This contradiction became an ongoing joke in future meetings. On retrospective reflection, Smit speculated that his reliance on ChatGPT's approval might stem from feelings of impostor syndrome triggered by the AI's challenge to their expertise. This sentiment is likely exacerbated by OpenAI using standardized examination scores to evaluate the performance of language models like ChatGPT [44]. All three researchers in this study scored less than GPT-4 on Graduate Record Examination (GRE)—a prerequisite (at the time) for applying to graduate schools in the U.S. Seeking ChatGPT's approval perhaps is merely reifying society's implicit (or explicit) tendencies to assess human value using numbers (e.g., [56]). It is not impossible to imagine a future where ChatGPT's h-index would be higher than all human researchers resulting in academics confronting similar insecurities.

Another emotion frequently mentioned in reflections and weekly meetings was frustration. Unlike embarrassment, frustration is a commonly experienced and widely studied emotion in HCI [24]. Frustration is experienced when a computing system prevents users from attaining their goals [24]. Tanusree experienced frustration when she was trying to perform an image description task, and ChatGPT kept generating nonsensical descriptions. Similarly, Pratyasha found ChatGPT's spurious response to "literature on digital sex work in Bangladesh" frustrating and misleading. In both cases, the frustration was intensified when ChatGPT failed to take responsibility for its inaccuracies. Its default response, "I am an AI language model…I cannot guarantee the accuracy or relevance of the information provided," is insufficient, as Smit pointed out in the third weekly meeting, "When it's giving you information, there are no such disclaimers. But when you call it out, it quickly falls back and apologizes." This incoherent shift from an omniscient LLM to just an LLM is the most frustrating part of using ChatGPT. The natural tendency of all three researchers in these scenarios was to argue about the veracity of the responses (exemplified in §4.2). However, these interactions are often pointless and highlight the inscrutable computational infrastructure that users depend on, leading to feelings of alienation and strangeness [50]. Smit explained this feeling using the existential concept of "absurd" [7], emphasizing the frustration of using and arguing with ChatGPT.

Although LLMs do not have real emotions, they have abilities to evoke emotions. The consequences of using a system with such abilities could be extreme—evidenced by a man reportedly committing suicide after interactions with an emotive chatbot based on an open-source alternative to the GPT model [59]. The appropriateness of using a "humanness" metaphor [14, 46] as a design tool is a topic of great interest in the CUI community. However, the implications of this deliberate design decision are manifesting expeditiously with LLMs. Even if ChatGPT never identifies itself as a human and denies having the ability to think, feel, or judge, it heightens its social presence using deceptive patterns [37], such as using the pronoun "I" and mimicking the effect of a person typing. However, these patterns can be offset by designing interfaces that are more honest about their capabilities and reduced reliance on anthropomorphism. One effective strategy is incorporating confidence scores, commonly used in AI-assisted decision-making [60]. While such approaches could compromise the glitzy human aspect of ChatGPT, they would lead to a more realistic and reliable use of LLMs, serving as a sounding board rather than an expert.

**4.2 A Self-Aggrandizing, Loveable Rogue AI that Just Can't Admit When It's Wrong (and apologizes profusely)**

*Tanusree: I decided to use ChatGPT as a search engine for literature curation. Though the list of papers curated by ChatGPT seemed legit, my initial excitement turned into frustration when I found eight of ten papers nowhere to be found on popular databases and even had either incorrect titles or authors. For example, ChatGPT suggested that the paper "Deep Convolutional Neural Networks for Image Classification: A Comprehensive Review" was by S. Kiranyaz et al. (2019) (hallucinated result), where authors name was incorrect. It even provided me with a paragraph that was supposed to be the abstract and an invalid URL for this elusive paper. I tried to give ChatGPT the benefit of the doubt by asking, "if the paper was legit." It kept feeding me with incorrect responses, even after I gave a hint that "author names and title paper don't match." This back-and-forth continues seven times! It was like talking to a stubborn toddler who refused to accept that the sky was blue. Finally, I provided information that "authors are Rawat, W., & Wang, Z. (2017)", instantly ChatGPT, being the humble AI, apologized and said, "You are right."*

These events evoked a "sentimental" reaction that conveyed a negative connotation [19], compelling Tanusree to elicit an admission of guilt from ChatGPT. In addition, the first few mistakes shaped Tanusree with a negative cognitive and emotional trust towards ChatGPT [21] which diverges from previous literature that posits a low level of trust trajectory followed by a positive development over time [54]. Tanusree was determined to hold ChatGPT accountable and coerce it to utter the phrase "I was wrong" [29]. It further complicates Tanusree's feelings about whether she was seeking empathy (make ChatGPT understand her intention) or accountability [51]. Whereas conceptualizing computer systems as autonomous moral (accountable) agents are debatable; thus, ChatGPT cannot be deemed entirely autonomous (thus, fully accountable) as it operates under the direction of external forces (such as algorithms and data fed into it) [58]. Moreover, ChatGPT's abrupt transition to admitting mistakes on the eighth attempt signaled two concerning indications—(1) its inclination to prioritize users' happiness by providing information, even if inaccurate; and (2) the transition was suggestive of a shift from an unyielding stance to a heightened susceptibility to persuasion, similar to that of a naive child. This parallel was also evident in Pratyasha's experience, highlighting ChatGPT's inadequacy in curating literature for sex workers in the Global South, where ChatGPT primarily directs resources through a moralist lens, potentially leading to bias. Similarly, Smit found ChatGPT's alt-text generation for box plot figures unsatisfactory, as it offered good tips on how to do it but failed to deliver in practice as Smit said: "It talks the talk but does not walk the walk."

These experiences highlight the importance of accountability in the development of language models such as ChatGPT, and the need for responsible AI thinking [1]. It raises questions about what the future of language models will look like, especially when, in our case, ChatGPT is inclined to exhibit contrition and amend its prior response when confronted with even a modicum of evidence that "you were incorrect, and this is the correct answer" as Tanusree experienced. Correspondingly, Smit perceives ChatGPT as a potential co-writer, whereas numerous entities have abruptly integrated ChatGPT as a constituent of their existing products. Such considerations raise issues of whether the future constraints ought to be treated as a rubric of legitimation or deception. Furthermore, there could be multiple futures for various stakeholders with design and legitimacy constraints [3], as Garcia's temporal model has shown that designing without considering the future is impossible [20]. In the next 10 years, if we were to adopt ChatGPT in personalization for blind users where they confirm the absence of nudity in an image before posting it on social which fails to identify correctly, it could lead to psychological distress as Tanusree found a completely incorrect image description of a picture within ChatGPT.

The ecological survival of human-nonhuman relations requires resilient modes of framing, particularly when used for vulnerable groups (i.e., blind users), political and unrepresentative values exploration, etc. The Futurama exhibit at the 1939 New York World's Fair is an example of how a utopian image of the future (America in 25 years) can eventually impact society in unforeseen ways. In particular, Bel Geddes's General Motors' internal combustion engine failed to account for complex societal consequences, including insurance fraud, decline of automobile-dependent cities in addition to environmental pollution, road rage, and accidents [61]. Although the possibilities that ChatGPT offers may be exhilarating, it is imperative that we consider the potential points of failure and the measures we can take to minimize risks, which are the foundation of accountability and responsible AI thinking. Moreover, it is essential to increase awareness of how different stakeholders negotiate the future as a resource for various purposes as well as embrace multiple futures and design with resilience.

### 4.3  Situated XAI: The Odyssey of the Marginalized and the Global South

*Pratyasha: "We've detected suspicious behavior from phone numbers similar to yours. Please try again later or contact us through our help center at help.openai.com." I got five of these warnings when I tried to sign up for ChatGPT with my local phone number. I attempted a few more times before I decided to try again with my mother's contact, but it kept showing the same alerts. I was unable to create an account despite repeated attempts, wondering if I could make a contribution to the paper I had intended to collaborate on. As the last option, I considered using my sibling's UK phone number to register. He gave me the OTP, and to my surprise, I was in within seconds!*

In contrast to Smit and Tanusree, Pratyasha's initial interaction with ChatGPT was marked by a unique concern regarding the platform's potential for harboring xenophobic tendencies. Considering the AI field's focus on research, development, and design is predominantly centered on the "West" or "Global North" [47], AI systems such as ChatGPT have a disproportionate effect on the marginalized segments of the society [6]. Pratyasha was concerned about being stigmatized as "suspicious" due to her geographical origin, particularly when juxtaposed with a contact from the West. This unique experience was further amplified in the additional usage of ChatGPT within the larger context of Global South when the instances of transgressions against Western cultural norms pertaining to slavery and discrimination were flagged with a red-box warning; however, the reference to the Brahmin caste as the superior, and Shudra as "Chamar" (an extremely offensive term for menial workers in this region) did not elicit any warning regarding policy violations. Even the short biography provided by ChatGPT on world leaders included critiques of Modi, Hasina, or Xi Jinping, yet no criticism was attributed to Biden. In addition to its innate bias against individuals from diverse strata, ChatGPT provoked bias in regular conversations fostering additional prejudicial views regarding gender like other AI systems [31]. While Pratyasha's language holds a gender-neutral pronoun for individuals, the translation service by GPT assigned gender to specific tasks within the translated sentences. (Ex: "He reads," "He earns money," "She cooks," and "She cleans the house", etc.) Social stereotypes and discrimination may be the outcome when the data used to train a language model contains biased representations of particular groups of people. For marginalized groups, this lack of fairness can prevent them from trusting these models and result in inaccurate or biased predictions about such populations [32,57]. The flawed stance demonstrated by ChatGPT towards content pertaining to the context of the South was further substantiated by the false and misleading classification of sex work as "illegal" in Pratyasha's country while searching for relevant literature on digital sex work. For both research and design purposes within the context of this region, Pratyasha encountered difficulties in relying on ChatGPT without manual verification. Introducing transparency, and XAI here could provide end users explanations to foster greater interaction by enabling them to act out and retrace AI/ML results, for instance, to check the correctness of the results [26]. The more critical use of ChatGPT revealed crucial concerns about how the society's most vulnerable, impoverished, and underprivileged groups can suffer unfavorably from AI systems [6, 18]. In a comprehensive dialogue with ChatGPT regarding the design of a platform cooperative [49] that is democratically governed and caters to the needs of marginalized domestic workers in Pratyasha's locality, the algorithm consistently advocates for a business model similar to gig job platforms, claiming to empower workers through secure employment opportunities. This model, managed by third parties instead of the workers themselves, has been proven to be facilitating severe exploitation and abuse of the intended beneficiaries [2, 23, 25], which helps reinforce the capitalist ideology and stands in stark opposition to the notion of "cooperative."

This provocation deliberately questions the fairness, transparency, and trustworthiness of ChatGPT in the context of Global South, and advocates for an explainable system situated in the context of its users. The potential of AI to address significant issues in the Global South, such as healthcare, poverty, agriculture, education, and other high-stakes sectors, has attracted increasing interest from governments, businesses, and academia in recent years [4, 22, 38, 39, 40, 52]. Evidently, AI has been known to aggravate and promote prevalent social issues like bias and prejudice [5, 41]. Therefore, it is important to take steps to ensure that AI systems are accessible and easy to comprehend for the individuals who will use them, especially those from underrepresented groups. Since scale and complexity are what currently enable successful AI, the inevitable calls for transparency led by the Black Box problem will be difficult to address. Ensuring explainability could be considered in this regard; nonetheless, this is terribly understudied how it can be efficient for the marginalized groups in contrast to the relatively techno-literate users of the Global North [43]. The integration of XAI systems in the Global South countries can be beset by the detrimental reflection of contextually inaccurate data and a lack of situated awareness during implementation, which may result in misplaced trust and over-estimation of AI capabilities [17, 43]. Here is where the term "situated XAI" [16] can come into play, concentrating on what explainability, autonomy, control, and trust essentially mean to individuals from diverse backgrounds instead of looking at how people interact with technologies. The establishment of an AI global governance entity [11] with scrupulous attention to the context of Global South is vital for effectively addressing the social, economic, and political disruptions that exceed the purview of individual governments, corporations, and academic or civil society groups. Instead of following the North-centric notion of XAI, in order to understand how explainability can be fostered more effectively within these communities, we advocate for further research and study into the implications of socially situated XAI.

## 5 CONCLUSION

In this paper, we use a month-long trioethnography to reflect on the use of ChatGPT in HCI research. Using our collective experiences, we reflect on (1) the impact of ChatGPT on our emotional states with an emphasis on frustration and embarrassment, (2) the lack of accountability and consideration of the future as a design rubric and raise (3) questions about bias from a Global South perspective. We hope these provocations serve as a call to action for the CUI community and direct focus on the need for further research, including design guidelines and governance, to address the diverse range of uses and users in this rapidly expanding field.

[62] Introducing ChatGPT. Retrieved from https://openai.com/blog/chatgpt